# Optics undergraduate experiments using smart (and not so smart) phones

V. L. Díaz-Melián[1], L. A. Rodríguez[2], F. Pedroso-Camejo[2], J. Mieres[2], Y. de Armas[3], A. J. Batista-Leyva[4],

and E. Altshuler[1]

[1]Group of Complex Systems and Statistical Physics, Physics Faculty, University of Havana, 10400 Havana, Cuba

[2]Pedagogical Science University "Enrique J. Varona" Havana, Cuba (ealtshuler@fisica.uh.cu)

[3]Institute of Cybernetics, Mathematics and Physics (ICIMAF), 10400 Havana, Cuba

[4]Instituto Superior de Ciencia y Tecnología Aplicadas (InSTEC), University of Havana, 10400 Havana, Cuba

**Abstract**

Smartphones may be seen as miniature toolboxs to perform Physics experiments. In this paper, we present three different "optics workbenches" mainly based on the light meter of a smartphone. One is aimed at the precise study of Malus law and other effects associated to linearly polarized light, the second allows quantifying the light intensity distribution of diffraction or interference patterns projected on a screen, and the third demonstrates the so-called inverse square law obeyed by the light from a pointlike source. These experiments allow to quantitatively demonstrate three fundamental laws of optics using quite inexpensive equipment.

I. INTRODUCTION

Quite probably, the sensors typically found in smart-phones {light meters, proximity sensors, accelerometers, gyroscopes, magnetometers and even barometers were not included by their designers for educational or scientific purposes. But the fact is that thanks to their existence, smartphones constitute a truly portable tool-box for Physics experiments. In fact, they have been used to analyze free falling objects [1], oscillations of coupled springs [2], the magnetic field of small magnets [3], rolling motion [4], a simple pendulum [5] constant motion [6,7] and optical phenomena [8,9] among a number of other scenarios of interest for the Physics undergraduate lab. Furthermore, smartphones are also compact and relatively heavy-duty devices than could play a role in certain experiments beyond Physics teaching: from the measurement of effective gravities [10], to the characterization of coils and magnets used in materials science research at moderate magnetic fields [11].

In this contribution, we briefly introduce three different applications for Physics laboratories aimed at life science undergraduate students, based on the light meters of smart phones. The proposals can be expanded to advanced teaching, if more complex operations and data analysis are involved. The experiments shown here were conceived and tested as final projects of the one-semester course "Advanced Physics Experiments" (Experimentos Avanzados de Física", EXAV) offered at the Physics Faculty, University of Havana.

II. EXPERIMENTAL SETUPS AND RESULTS

A. Workbench for polarized light studies

The first proposal is a workbench for polarized light studies, depicted in Fig. 1(a). It consists in an inexpensive wooden "optical bench" with a 12V LED light source, two polaroids identied as "polarizer" and "analyzer" in the figure, and an adjustable holder to attach smart phone upside down, in such a way that the light meter sensor (typically near the "head" of the apparatus) faces the analyzer. The light meter of the smart phone can be easily accessed by using the Physics Toolbox Sensor Suite application [12], and the numerical values of the illuminance (in lux) can be read from the phone screen with relative ease.

The non-polarized light produced by the LEDs is linearly polarized by the polarizer, then passes through the analyzer, and is detected by the lux-meter of the smart-phone. The analyzer can be rotated in order to change the angle $\theta$ between its polarizing axis and that of the polarizer. The relation between the irradiance reaching the smartphone $I$ and that emerging from the polarizer, $I_0$ is supposed to follow Malus law [13]:

$$I = I_0 \cos^2 \theta \tag{1}$$

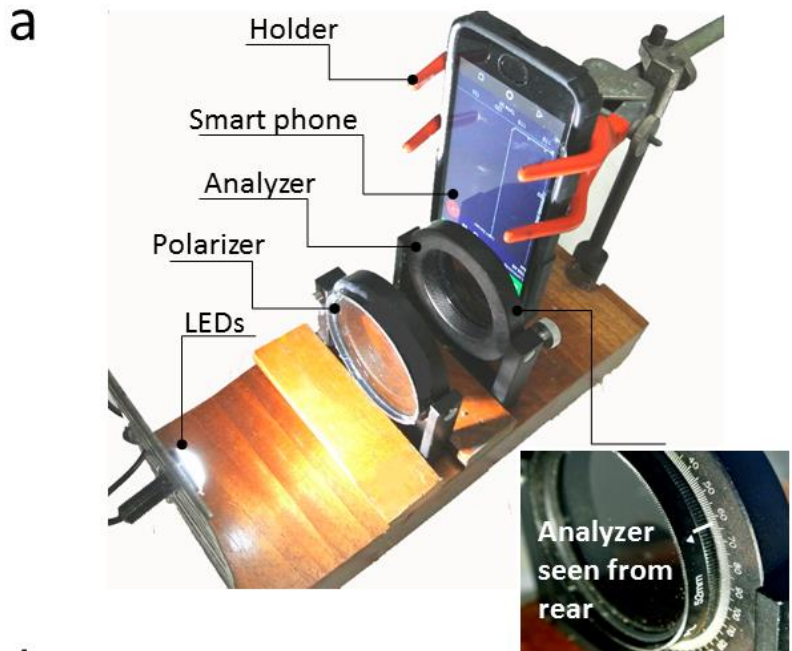

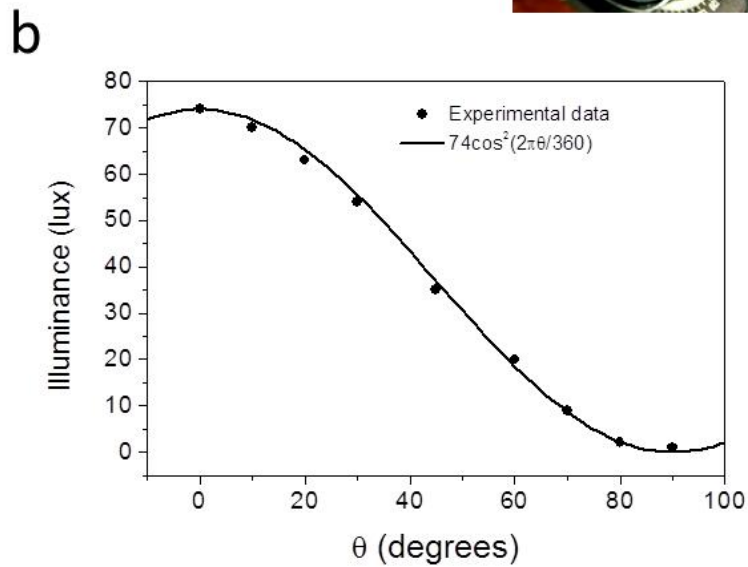

FIG. 1: Quantifying Malus law. (a)A home-made polarized light bench. In the picture, the smartphone is upside-down, since the light meter is typically at its \head". (b) Fit of Malus law (continuous line) to the experimental data (circles).

Fig. 1(b) shows that equation (1) nicely follows the experimental points measured with the help of a Moto smartphone, thus demonstrating the validity of Malus law (the quality of the curve may suffer if very low-quality polaroids are used). It should be noticed that the smart phone's light sensor displays illuminance [14] (which is a subjective measure of irradiance hitting a unit surface [15]), but its subjective nature does not influence our experiments, so we have written just "irradiance" in the formulas [16].

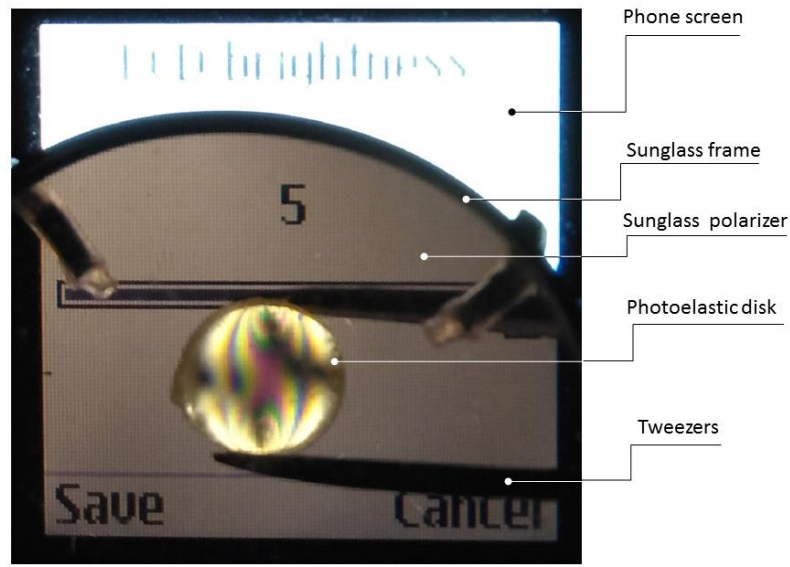

FIG. 2: A minimalistic polarization workbench. From the back, an old cell phone illuminates with linearly polarized light a photoelastic disk slightly compressed by tweezers. The disk is sandwiched between the cell phone and a sunglass with polarizing direction at a right angle relative to the phone's po-larization direction. The picture has been taken with a smart-phone's camera. The light pattern seen on the photoelastic disk illustrates the complex stress distribution inside it.

The setup can be also used to perform other experiments, such as checking out that screens from older cell phones (i.e., those without a depolarizing sheet) produce linearly polarized light. This can be done by turning off the LEDs of our set up, and substituting the polarizer by the cell phone to be analyzed, with the screen facing the analyzer. Then, Malus law can be reproduced by rotating the analyzer, demonstrating that the light emitted by the phone screen is linearly polarized. This possibility has been previously reported by Monteiro and coworkers, in a paper we came across after writing the first version of this manuscript [8].

Taking that into account, it is possible to set up a truly "minimalistic" bench for polarized light experiments. It looks evident, for example, that the bench illustrated in Fig. 1 can be reduced by removing both the LEDs and the polarizer, and substituting it by a "not-so-smart" phone (or also by a screen of an old laptop, among other possibilities). Fig. 2, shows an example of an "extreme" minimalistic bench. There, we show a picture taken with a Moto smartphone of a photoeslastic disk sandwiched between an old Samsung "not-so-smart" phone, and a polarized sunglass. The Samsung phone provides linearly polarized light from the back, whose polarization direction is tilted by the optical activity associated to the local stresses "felt" by the photoelastic element due to compression: the resulting pattern is revealed by the sunglass playing the role of an analyzer, whose polarizing axis is almost at a right angle relative to that of the Samsung phone screen. This experiment can be taken as a motivational illustration at the introductory lab level, but, if the photoelastic effect is quantified and modeled, it can be moved to the advanced curriculum.

**B. Non-goniometric quantication of interference and diffraction**

The second setup allows quantifying the interference or diffraction patterns resulting when laser light goes through different obstacles. We will illustrate here the diffraction pattern resulting from a green laser beam passing through a single vertical slit of horizontal width a. Differently from conventional "goniometric measurements", here we use the lux meter of a smartphone to determine the irradiance of the pattern projected on a screen attached to the phone, which is located at distance $L$ $(L >> a)$ from the slit. The phone rests on a teaching-grade optical bench oriented perpendicularly to the laser beam, in such a way that it can be manually slid horizontally, as illustrated in Fig. 3(a). A clear cardboard screen is taped to the phone in such a way that a hole made with an office hole punch allows the light to enter the light sensor, while the pattern is displayed on the cardboard surface. Then, the intensity of the diffraction pattern can be measured every few millimeters (we notice that the horizontality of the phone path can be checked using other of its sensors: the g-force meter!). Fig. 3(a) shows that a diffraction pattern obtained with a green laser can be visualized easily up to second order. If we assume that the phone scans the pattern along the $x$ direction, the irrradiance should follow the formula [13]:

$$I = I_m \left( \frac{\sin \xi x}{\xi x} \right)^2 \qquad (2)$$

where $I_m$ is the maximum irradiance at the screen, and

$$\xi = \frac{\pi a}{\lambda L} \qquad (3)$$

Fig. 3(b) shows the good fit of the above formulas to the illuminance record of an actual diffraction experiment. Notice that, as the light in mono chromatic, the illuminance is proportional to the irradiance.

**C. The inverse-squared-distance law**

The third setup aims at the experimental confirmation of the inverse-square proportionality law followed by the irradiance (and related magnitudes) emitted by a pointlike light source with distance.

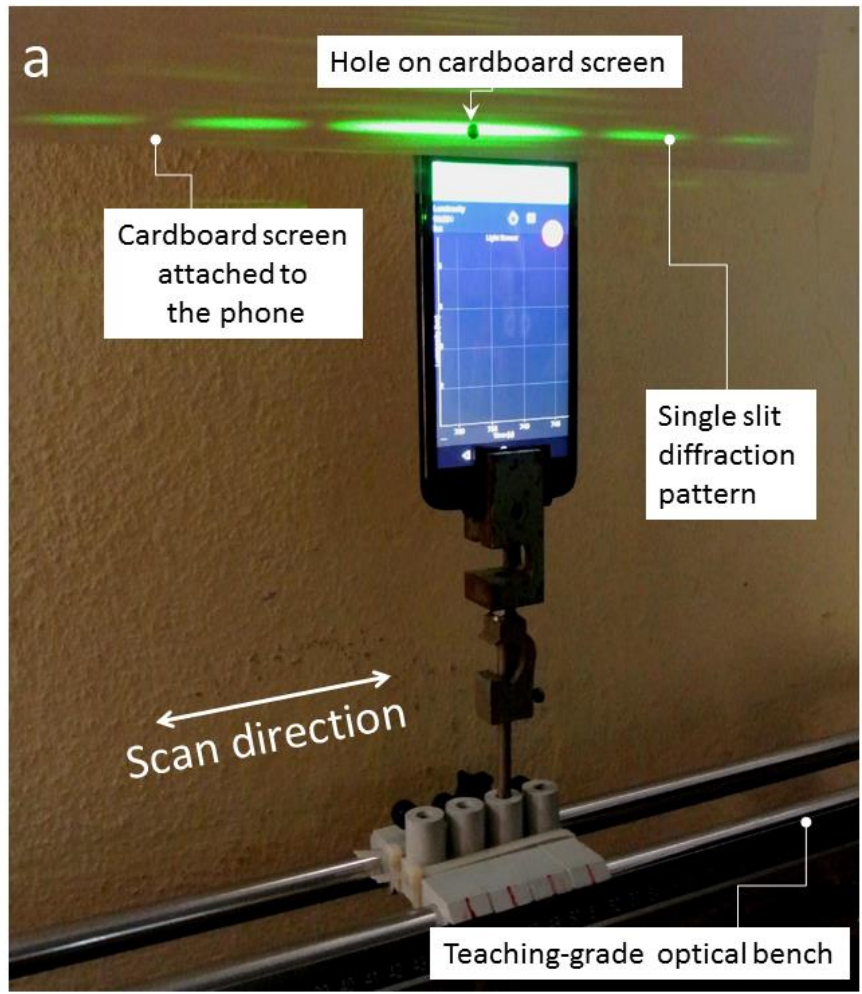
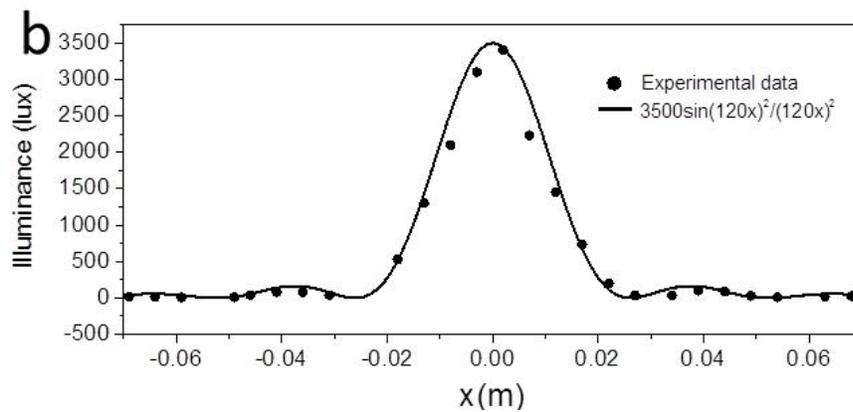

FIG. 3: Scanning diffraction patterns (a) Photograph of a diffraction pattern being scanned by a smart phone (b) Illuminance recorded along a diffraction pattern projected by a 140 micrometers width slit, 7m away from the phone, using a green laser of wavelength 532 nm (the pattern shown in (a) was obtained with slightly different parameters).

On one extreme of a teaching-grade optical bench like the one shown in Fig. 3(a) we have fixed a white LED. A smart phone slides facing the LED along the optical bench, thanks to a holder similar to the one shown in Fig. 3(a), (but the phone has been rotated in such a way that it faces the LED source). Due to its simplicity, we are not showing a picture of the setup; we just illustrate one set of experimental data and its t to the inverse-squared-distance law in Fig. 4. The fit clearly demonstrates the validity of the law. It is worth noticing that smart phones have been used very recently to characterize in a similar way linear light sources [9].

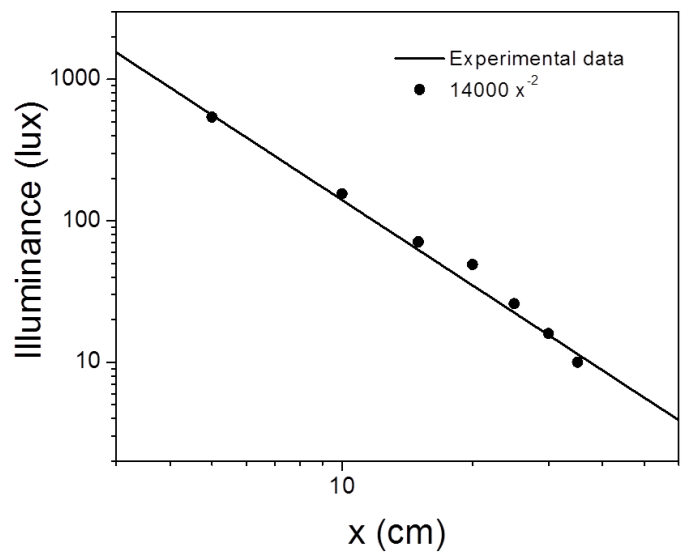

FIG. 4: Fit of the inverse-squared-distance law (continuous line) to the experimental data (circles)

III. CONCLUSIONS

Smartphones have been quite extensively used in introductory Mechanics experiments, involving not only the phone's accelerometer and gyroscope, but also its magnetometer and light sensor. In this paper we use smart-phones to tackle a different field of Physics, where much less work has been published: Optics.

We demonstrate that the light sensor of a smartphone can be used to design and construct very simple and inexpensive setups allowing the quantification of fundamental optical phenomena: the Malus formula, diffraction (and interference) through slits, and the inverse-squared-distance formula. In addition, we propose the use of the screen of a second phone or laptop as an inexpensive source of linearly polarized light, resulting in extremely inexpensive and simple setups. A key simplifying element of our setups is that they do not involve any goniometers.

Extending the span of experiments covered by our work benches is not difficult: the polarization setup can be used to measure the optical activity of certain solutions, while the inverse-squared-distance one can be used to quantify the absorption of light by a substance, to put two examples.

**Acknowledgements**

We acknowledge A. C. Martí for calling our attention to references [8] and [9] after submitting the first version of the present manuscript. A. Pentón-Madrigal is gratefully thanked for advice about the equipment availability and, in particular, for providing a variable slit worth of being on display in a museum. E. Altshuler found inspiration in the late M. Álvarez-Ponte.

[12] https://play.google.com/store/apps/details?id=com.chrystianvieyra.physicstoolboxsuite

[13] Hetch E Optics, 5th edition. Pearson, 2016

[14] Landsberg G S Óptica vol. I, Mir 1983

[15] The illuminance is the luminous flux per unit surface perpendicular to the flux. The luminous flux is the flux of radiant energy (or radiant flux) weighted by the photopic spectral luminous efficiency. So, different sources having equal radiant flux could have different luminous flux, provided they emit at different wavelengths

[16] Let us suppose that a non-polarized polychromatic light beam passes through a polarizer-analyzer system. When passing through the polarizer, each wavelength diminishes its irradiance in a fixed percent (50% for an ideal polarizer). When passing through the analyzer, a fixed amount, proportional to the square cosines of the angle $\theta$ is eliminated for each wavelength. So, from the analyzer exits an irradiance equal to the sum of its value for every wavelength. But the illuminance for each wavelength is proportional to its irradiance, so the overall illuminance will be proportional to the overall irradiance: all in all, Malus law also holds for polychromatic light beams